\documentclass{JHEP3}

\usepackage{amsmath,graphics,epsfig,amsthm}

\title{Explicit Formulas for the Scalar Modes in Seiberg--Witten Theory with 
       an Application to the Argyres--Douglas Point}

\author{Nikolas Akerblom and Michael Flohr\thanks{On leave of absence from 
          ITP, Universit\"at Hannover,
	  Appelstra\ss e 2, 30167 Hannover, Germany}\\
	ITP, Universit\"at Hannover, Appelstra\ss e 2, 30167 Hannover, Germany\\
	Physikalisches Institut, Universit\"at Bonn, Nu\ss allee 12, 53115 Bonn,
          Germany\\
	E-mail: \email{akerblom@itp.uni-hannover.de}, 
                \email{flohr@itp.uni-hannover.de}}
	
\abstract{General formulas for the scalar modes $a_i$ and $a^i_D$ in the 
          Seiberg--Witten $SU(N)$ setting are derived in the cases with and 
          without massive hypermultiplets. Subsequently these formulas are 
          applied in a study of the $SU(3)$ Argyres--Douglas point.
          We use this example to study the question, whether the scalar modes
          admit an interpretation in terms of BPS mass states everywhere in
          moduli space.
          The paper collects, in an appendix, various facts on the function Lauricella $F_D^{(n)}$,
          which naturally appears in the derived formulas.}


\preprint{}

\begin{document}

\section{Introduction}

In a much celebrated work \cite{Seiberg:1994rs}, 
Seiberg and Witten found an exact solution
to $N=2$ supersymmetric four-dimensional Yang--Mills theory with gauge group
$SU(2)$. This paper initiated a whole new, tree-sized, branch of research
leading to a vast set of exactly solvable Yang--Mills theories
in various dimensions and with various degrees of supersymmetry.  For some
basic or introductory works see, for example, 
\cite{SW5,SW6,Klemm:1994qs,Argyres:1994xh,Hanany:1995na,Lerche:1996xu, 
D'Hoker:1999ft},
and references therein. Of
particular interest for these solutions is the understanding of the moduli
space of vacua, which in many cases turns out to be a hyperelliptic
Riemann surface. In particular, simply-laced Lie groups lead to spectral
curves which are hyperelliptic.

The BPS spectrum of such a model is entirely determined by the periods of
a special meromorphic differential 1-form on this Riemann surface, the
Seiberg--Witten differential $\lambda_{{\rm SW}}$. A general hyperelliptic
Riemann surface $\Sigma$ can be described in terms of two variables $w,Z$ in the
polynomial form 
\begin{equation}\label{eq:surf}
    w^2 + 2A(Z)w + B(Z) = 0
\end{equation}
with $A(Z),B(Z)\in\mathbb{C}[Z]$. After
a simple coordinate transformation in $y=w-A(Z)$, this takes on the 
more familiar form $y^2 = A(Z)^2 - B(Z)$. But we might also write
the hyperelliptic curve in terms of a rational map if we divide the
defining equation \ref{eq:surf} by $A(Z)^2$ and put $\tilde{w}=w/A(Z)+1$
to arrive at the representation
\begin{equation}\label{eq:ratmap}
    (1-\tilde{w})(1+\tilde{w}) = \frac{B(Z)}{A(Z)^2}\,.
\end{equation}
This form is very appropriate in the frame of Seiberg--Witten models, since
the Seiberg--Witten differential can be read off directly: The rational map
$R(Z)=B(Z)/A(Z)^2$ is singular at the zeroes of $B(Z)$ and $A(Z)$, and is
degenerate whenever its Wronskian $W(R) \equiv W(A(Z)^2,B(Z)) =
(\partial_ZA(Z)^2)B(Z) - A(Z)^2(\partial_ZB(Z))$ vanishes. This is precisely
the information encoded in $\lambda_{{\rm SW}}$ which for arbitrary
hyperelliptic curves, given by a rational map $R(Z)=B(Z)/A(Z)^2$, can be
expressed as
\begin{equation}\label{eq:lamSW}
    \lambda_{{ \rm SW}} = 
    % -Z\frac\left\{ \rm d(arctanh}\sqrt{1-R(Z)})}{\pi i} = 
    \frac{Z}{2\pi{\rm i}}\,{\rm d}(\log\frac{1-\tilde{w}}{1+\tilde{w}}) =
    \frac{1}{2\pi{\rm i}}{\rm d}(\log R(Z))\frac{Z}{\tilde{w}}
    = \frac{1}{2\pi{\rm i}}
    \frac{W(A(Z)^2,B(Z))}{A(Z)B(Z)}\frac{Z\,{\rm d}Z}{y}\,.
\end{equation}
Note that the denominator polynomial being a square guarantees
the curve to be hyperelliptic. It is this local form of the Seiberg--Witten
differential which serves as a metric ${\rm d}s^2 = |\lambda_{{\rm SW}}|^2$ 
on the Riemann surface. And it is this local form which arises as the 
tension of self-dual strings coming from 3-branes in type II string theory
compactifications on Calabi--Yau threefolds.

The crucial point in Seiberg--Witten theory is that the physically relevant
information of the BPS mass spectrum is encoded in the periods of the above
defined $1$-form. In general, the hyperelliptic curve has $2N_c$ branch
points and is thus of genus $g=N_c-1$. Any element $\gamma$ of the homology of
the curve defines a period and hence the mass of a BPS state,
\begin{equation}
  Z_\gamma = \oint_\gamma \lambda_{{\rm SW}}\,,\ \ \ \ m_\gamma = |Z_\gamma|\,.
\end{equation}
If $N_f$ (massive) hypermultiplets are present, the homology basis gets 
extended by $N_f$ loops encircling each of the differential's poles in order to pick up
residue terms, if the contour $\gamma$ is deformed in such a way that it
crosses a pole.

The appealing picture emerging from the analysis of the Seiberg--Witten solution
of supersymmetric low-energy effective field theories is that the geometry
of an auxiliary Riemann surface $\Sigma$ entirely determines the physics. Given
the surface and a special $1$-form on it, the properties of particles are
encoded in the periods of this particular $1$-form, the Seiberg--Witten
differential. Choosing a canonical homology basis $\{\alpha_i,\beta^i\}$ with
$1\leq i<N_c$, such that $\alpha_i\cap\alpha_j=\beta^i\cap\beta^j=0$ and
$\alpha_i\cap\beta^j=\delta_i^{\phantom{i}j}$, defines electric-like and
magnetic-like quantum numbers for particles. More precisely, every
$\gamma\in H_1(\Sigma,\mathbb{Z})$ can be written as a
linear combination with integer coefficients,
\begin{equation}
  \gamma = \sum_i q^i\alpha_i + g_i\beta^i\,,
\end{equation}
defining in turn dual pairs of quantum numbers $(q^i,g_i)$. Stable particles
are those, for which all these pairs consist of coprime numbers. As we see, the
definition of `electric' and `magnetic' quantum numbers is fixed by the
choice of a basis in the homology of $\Sigma$. It is generically independent
of small variations of the location of the branch points of the Riemann
surface, i.e.\ the basis is generically invariant under small variations of
the moduli of the surface. However, there is typically no single canonical
homology basis, which can be used for all possible variations of the moduli.
This will concern us in some detail later on. This is an important point,
since varying the moduli usually implies a deformation of the contours,
which encircle, once and for all, chosen pairs of branch points. However,
since the Seiberg--Witten differential has zeroes and poles, such 
deformations are not always possible without crossing one of these.

The approach taken in this paper emphasizes the geometrical picture for
the BPS states, namely that they are described in terms of period integrals
taken along homology contours. In this picture, one immediately sees
which BPS states can become massless under what circumstances. This is
especially easy to see, when the Riemann surface is realized as a branched
covering of the complex plane (or sphere). As soon as two branch points
coalesce, the BPS state corresponding to the cycle
$\gamma_{ij}$, which encloses precisely the two branch points $e_i$ and $e_j$ and none
of the poles of the Seiberg--Witten differential, will become
massless, since $Z_{\gamma_{ij}}\rightarrow 0$ for $\gamma_{ij}\rightarrow 0$,
i.e.\ for $\lvert e_i-e_j \rvert \rightarrow 0$ under a suitable variation of the moduli.
Another important advantage of our approach is that we do not need to find
Picard--Fuchs equations, which is a very hard task for gauge groups of
rank greater than two.

\subsection{Outline of motivation and strategy}

Up to here, we have not really deviated from the common point of view.
However, we would like to emphasize the following point in this work:
Even if it is clear that the periods $a_\gamma(u)$ of the Seiberg--Witten 
differential $\lambda_{{\rm SW}}(u)$ are complex analytic functions of the 
moduli $u_k$ of the Riemann surface $\Sigma$, it appears to us that the 
interpretation of a given $a_\gamma(u)$ as describing BPS states for all
of the moduli space spanned by the $u_k$ may break down. The reason is
hidden in the complicated and highly non-trivial properties of the
functions $a_\gamma(u)$ under analytic continuation. Let us start with a
given $a_\gamma(u)$ for such values of the $u_k$ that it is perfectly well
defined as a period integral along the cycle $\gamma$. Then it has an
asymptotic expansion in the $u_k$. Now we seek its analytic continuation
for other patches in the moduli space of the $u_k$, which can be
done, for example, by studying the monodromy properties of $a_\gamma(u)$.
Of course, analytic continuations exist for any patch in the moduli
space. But what strikes us is that many of these seem not to have
an interpretation as simple period integrals anymore.

More precisely, we claim that the set of Seiberg--Witten periods,
given in terms of a basis of functions $a_i=a_{\alpha_i}(u)$ and
$a_D^j=a_{\beta^j}(u)$ for a suitable patch in the moduli space spanned
by the $u_k$, does not close under monodromy or analytic continuation.
This is to be contrasted with the well-known fact that, for example,
the standard Gaussian hypergeometric system, spanned by two solutions of
its defining second-order ordinary differential equation, is complete
under analytic continuation. This is precisely, what happens in the
simplest case of gauge group $SU(2)$. The resulting Riemann surface is
elliptic, and the Seiberg--Witten periods can be recast in a form given
in terms of ordinary hypergeometric functions. Thus, the theory can
be explored in all of moduli space, i.e.\ for all values of the single
complex modulus $u$. As we will show, higher rank gauge groups, which
lead to hyperelliptic surfaces, involve generalized hypergeometric
functions of several variables. In fact, we will have to deal with the
particular set of functions $F_D^{(n)}$, called Lauricella functions of type 
$D$. First of all, the defining differential equations are now partial 
differential equations, from which one might already guess that things get
worse. Indeed, not much is known about
analytic continuations of these functions. However, one thing is clear, 
namely that the set of the functions $F_D^{(n)}$ for a given $n$ is
not sufficient in order to define analytic continuations everywhere.
The problem is that one has to add further functions to this set,
a different sort of generalized hypergeometric functions $D_{p,q}^{(n)}$.
These functions seem not to have an integral representation which can
be interpreted as a period integral. 

Thus, our analysis raises the question whether analytic continuations of
Seiberg--Witten periods always possess an interpretation as central
charges for BPS mass states. Unfortunately, the complete answer to this
question is complicated by the fact that the map between the $N_c-1$ moduli
$u_k$, $k=1,\ldots,N_c-1$, and $\Lambda$ of the Riemann curve $\Sigma$ and
the $2N_c$ branch points $e_i$ is multi-valued and rather opaque.
A generic hyperelliptic surface of genus $g=N_c-1$ has a moduli space
of (complex) dimension $2g-1=2N_c-3$. This is clearly not exhausted
by the $N_c$ complex numbers $u_k$ and $\Lambda$ for $N_c>3$. For this
reason, will we concentrate mainly on the case of gauge group $SU(3)$,
since then the dimensions of the moduli space and the space spanned by
$u_1\equiv u$, $u_2\equiv v$ and $\Lambda$ match. Thus, the coordinates
$u,v,\Lambda$ may serve {\em locally\/} as coordinates of the moduli space.
For $N_c>3$, the vacuum expectation values $u_k$ and the cut-off $\Lambda$
will only sweep out a physically relevant sub-variety of the full
hyperelliptic moduli space. Our findings are nonetheless of a general
nature and hold true also for $N_c>3$. It is worth noting here that
adding of $N_f$ massive hypermultiplets, if we allow for complex masses
$m_r$, can make up for the missing moduli. In fact, $N_c-1$ vacuum
expectation values, $N_f\leq N_c$ masses and the cut-off $\Lambda$ precisely
sum to $2N_c$ complex numbers. It is well known that the theory
with $N_f=N_c$ massive hypermultiplets is conformally invariant, which
reflects the fact that conformal transformations can change three of the
$2N_c$ numbers to whatever we want\footnote{Actually, what can be transferred
by conformal transformations is the location of the branch points, and three
of these can be mapped to any desired values. However, the $2N_c$ branch points
are functions of the $N\leq 2N_c$ numbers $u_k$, $m_r$ and $\Lambda$. If
$N=2N_c$, we can locally invert these functions to obtain locally valid
coordinates in moduli space.}, leaving $2N_c-3$ free variables, which
is precisely the dimension of the hyperelliptic moduli space. 

As mentioned above, the case $SU(3)$ without hypermultiplets is the simplest
case we can consider. The phenomenon we want to study is, however, generally
valid. The Seiberg--Witten differential, expressed in
the branch points of the Riemann surface $\Sigma$, has a very rich structure
of singular sub-manifolds where special things may happen. The best known
case is that of two of the branch points flowing into each other. This causes
the associated BPS state to become massless. But as soon as the number of
branch points exceeds four, the possibility arises that more than two of
them might run into each other. In fact, we may consider $2g-1$ of the 
$2g+2$ branch points of a generic hyperelliptic surface to be free
variables, fixing the remaining three to, e.g.\ $0$, $1$, and 
$\infty$.\footnote{Which three we choose is up to us and can be adapted to
the situation at hand. Thus, in the following discussion we can assume
without loss of generality
that the singular point in question is of the form
$(0,\ldots,0,1,\ldots,1,\infty,\ldots,\infty)$.}
There are very many ways in which branch points can form
several subsets where all points in a subset are close to coinciding with each
other. Let us call the free variables $x_i$, $i=1,\ldots,2g-1$.
Typically, one first calculates the Seiberg--Witten periods in a patch of
moduli space where all $x_i$ are small. This patch corresponds to the
singular point $(0,0,\ldots,0) \in\mathbb{C}^{2g-1}$. Now, the number $M$ of
singular sub-manifolds intersecting at this point is given by the 
$2g-1$ possibilities $x_i=0$ together with the $\binom{2g-1}{2}$ possibilites
that $x_i=x_j$ for $i\neq j$. This makes in total $M=g(2g-1)$ possibilites.
We see that for gauge group $SU(2)$, i.e.\ $g=1$, this simply yields $M=1$
possibility, so nothing spectacular is expected to happen. But already for 
gauge group $SU(3)$, i.e.\ $g=2$, we get $M=6$ sub-manifolds, which intersect
at the singular point $(0,0,0)$. The general case is given by
singular points $(\underbrace{0,\ldots,0}_p,\underbrace{1,\ldots,1}_q,
\underbrace{\infty,\ldots,\infty}_{2g-1-p-q})$ or permutations thereof. 
%with $p$ zeroes, $q$ ones and $2g-1-p-q$ infinities. 
The number $M$ of intersecting sub-manifolds now is
\begin{equation}
  M = {\textstyle\frac12}p(p+1)+{\textstyle\frac12}q(q+1)+
  {\textstyle\frac12}(2g-1-p-q)(2g-p-q)\,.
\end{equation}
Considerable difficulties arise when one attempts to obtain a {\em complete\/}
set of solutions of any given hypergeometric partial differential system valid
in an entire neighbourhood of such a singular point, as soon as more than
$2g-1$ singular sub-manifolds intersect, i.e.\ as soon as $M\geq 2g$. In fact,
it appears that it is impossible to obtain a single fundamental set of
solutions valid in the entire neighbourhood of such a singular point. 
Instead, one can only construct sets of solutions valid in hypercones whose 
common vertex is the singular point in question. 

The hypergeometric system we will encounter here is the Lauricella system
$F_D^{(n)}$. It is worth noting the peculiar property that it is
impossible to construct any one fundamental set of solutions valid in the
whole neighbourhood of any possible singular points by the methods known
and employed here, whenever the number of free variables exceeds three.
Thus, we have a chance to deal with $SU(3)$ without hypermultiplets using
only the standard methods for analytic continuation. We will do so later,
and obtain in this way solutions valid in entire neighbourhoods of 
certain singular points. Already the case $SU(4)$ is out of reach for a 
complete treatment of the neighbourhood of even one singular point.

We remark here, that our approach works with a completely factorized form
of the Seiberg--Witten differential. Thus, our functions will formally depend
on the branch points, the poles (given by the masses of the hypermultiplets and
the point at infinity, where the Seiberg--Witten differential has a double pole)
and the zeroes of $\lambda_{{\rm SW}}$. Even though the zeroes are also
functions of the moduli and masses and are thus not independent of the
branch points and poles it is more convenient, and
also more economic, to treat the zeroes as additional variables. It does
not influence our argument later on, since the Seiberg--Witten period integrals
can always be decomposed by linearity into integrals depending solely on the
branch points and the poles. It is the number of truly independent variables
which counts. For gauge group $SU(2)$, we have just one variable, and hence
it is no surprise that everything can be reduced to ordinary
hypergeometric functions with their well-known analytic properties. For all
higher rank gauge groups, we have at least two independent variables.
The minimal case we can then run into is the Appell function $F_1$, a 
generalized hypergeometric function of two variables. It is known that
a complete set of solutions sufficing to construct analytical continuations 
in whole neighbourhoods around singular points consists out of 25 integrals,
only ten of these being of type $F_1$, the other 15 being Horn's functions
of type $G_2$. The latter do not possess integral representations in terms
of contour intergals along simple (Pochhammer) loops. In fact, double
circuit loops are needed as well, where each of the two loops encircles
two singular points. Thus, we arrive at the statement announced above that
the analytical continuation, althogh existent, does not everywhere admit
an interpretation in terms of periods of scalar modes yielding BPS mass
states. 

\subsection{Structure of the paper}

In the following, we will work out this line of argument in more detail,
concentrating mainly on one particular example, namely the Argyres--Douglas
$\mathbb{Z}_3$-point for the case of gauge group $SU(3)$, since this is
precisely such a singular point where several singular sub-manifolds
intersect.

In order to carry out this analysis we derive general formulas for the scalar
modes in Seiberg--Witten theory which involve the above-mentioned Lauricella
$F_D^{(n)}$. These formulas work directly in the picture of a branched
ramified covering of the complex plane and thus do not rely on the 
derivation and solutions of the associated system of Picard--Fuchs equations.

This paper is organized as follows: We begin by stating the necessary preliminaries from
`Seiberg--Witten theory.' Then we derive our formulas for the scalar modes with and without
hypermultiplets for gauge group $SU(N)$. As an application we consider Argyres--Douglas' $\mathbb{Z}_3$-point.
This is followed by a section ``Conclusions and Outlook,'' where we discuss
some possible consequences of our findings, suggesting directions of 
further investigation.

The paper closes with a rather long appendix on the Lauricella functions
$F_D^{(n)}$ which naturally appear in our derived formulas. The idea is to
collect all necessary results on these functions in order to make the paper
more self-contained.

\section{General formulas for the scalar modes}

In this section we derive our formulas for the scalar modes occuring in the
BPS mass formula. As was remarked in the introduction, one of our devices is the
factorization of the Seiberg--Witten differential so that the integrals involved in our
calculations can be identified as Lauricella functions of type $D$.

In the first subsection we gather the relevant expression for the Seiberg--Witten differential
and so on. Then, in the second subsection, we consider the $SU(N)$ case without massive
hypermultiplets, whereas in the third subsection we consider the case with
massive hypermultiplets.

For completeness, we mention that integration of the Seiberg--Witten form
can be achieved by other methods. First of all, the traditional way is
by setting up and solving the Picard--Fuchs system of the curve $\Sigma$ with
given form $\lambda_{{\rm SW}}$. Another method integrates the Seiberg--Witten
differential directly, after recasting it in modular coordinates with respect 
to the charge lattice generated by its periods \cite{geodesics1}.
Conformal invariance or isomonodromic properties of the periods can also
be used to find them in explicit form\cite{Cappelli,Flohr:1998ew}.
Finally, for certain cases such as $SU(2)$, functional properties of the
prepotential related to a certain kind of modular invariance can also be employed 
\cite{Matone:1995rx,Nahm:1996di}, see also \cite{Nekrasov:2002qd} for an
approach where the prepotential is computed directly, without determining the
scalar modes first.
Most of these methods become increasingly more difficult or cannot be
applied at all for higher rank gauge groups.

\subsection{The Seiberg--Witten differential}\label{ETC}

As is well known, the exact Seiberg--Witten solutions of SUSY Yang--Mills theories are obtained
like this: One writes down a certain hyperelliptic curve,\footnote{To be precise, there are
cases where the curve is not hyperelliptic. This happens for gauge groups which are
not simply-laced. However, we shall not consider those cases.} often called the spectral curve
of the theory, and the BPS spectrum for the particular theory is then determined by the periods
of a given meromorphic 1-form $\lambda_\mathrm{SW}$ on this curve.

For $N=2$ supersymmetric $SU(N)$ Yang--Mills theory without massive hypermultiplets the curve-defining
equation is (the $N$ is the $N$ in $SU(N)$)
\begin{equation}\label{curve}
y^2=A(x)^2-B\mbox{$\,:=\,$}
        \left(x^N-\sum_{k=2}^N u_k x^{N-k}\right)^2-\Lambda^{2N}
	=\prod_{i=1}^{2N}(x-e_i),
\end{equation}
and the Seiberg--Witten differential is
\begin{equation}\label{lambdasw}
\lambda_\mathrm{SW}=\frac{1}{2 \pi \mathrm{i}}
	\frac{\prod_{\ell=0}^{N-1} (x-z_\ell)}{\prod_{i=1}^{2N}\sqrt{x-e_i}}\,\mathrm{d}x,
\end{equation}
where the $z_\ell\;(\ell>0)$ are the zeros of $2 A'(x)B$, $z_0=0$ and the $e$'s are the zeros
of $y^2$. Usually the $e$'s are called \emph{branch points}.

This form of the Seiberg--Witten differential is not unique. We could add exact
terms to it which would not affect the computation of period integrals.
However, the form given here is singled out in string-theoretic derivations
of Seiberg--Witten low-energy effective field theories. There, the spectral
curve $\Sigma$ appears through intersecting branes, and the Seiberg--Witten 
differential is induced by the metric on $\Sigma$ which arises from the
tension of self-dual strings (in a derivation from type II string theory).

With these data one proceeds as follows: Let $\{\alpha_i,\beta^i\}_{1\leq i \leq N-1}$ denote a
canonical homology basis for our curve, i.e.\ one for which
$\alpha_i\cap\beta^j=\delta_i{}^j$, then the scalar modes are given by
$a_i=\int_{\alpha_i}\lambda_\mathrm{SW},\,a_D^j=\int_{\beta^j}\lambda_\mathrm{SW}$ and in
terms of these quantities the mass of a BPS state with charge $(\mathbf{q},\mathbf{g})$ is
given by $m_{(\mathbf{q},\mathbf{g})}\propto\lvert q^i a_i+g_ja_D^j\rvert$.
In a string-theoretic derivation, the homology cylces arise as projections from
three-cycles in Calabi--Yau compactifications. Other derivations of 
Seiberg--Witten theory exist, e.g.\ from $M$-theory, but they agree in the
canonical choice of $\lambda_{{\rm SW}}$ and its periods.

The case with massive hypermultiplets brings about some changes in the spectral curve as well as in
$\lambda_\mathrm{SW}$ which we shall communicate in due course (see Section \ref{massive} below).

In deriving our results, we will be making use of the following integral identity involving Lauricella
$F_D^{(n)}$; this function is considered more fully in the appendix as well as in \cite{Exton:yx}:
\begin{equation}\label{Euler}
\int_0^1 t^{a-1}(1-t)^{c-a-1}\prod_{i=1}^n(1-tx_i)^{-b_i}\, \mathrm{d}t=
	\frac{\Gamma(a)\Gamma(c-a)}{\Gamma(c)}\,F_D^{(n)}(a,b_1,\ldots,b_n;c;x_1,\ldots,x_n).
\end{equation}

Briefly, $F_D^{(n)}$ is a function of several (namely $n$) complex variables, defined by the
power series
\begin{multline}\label{pow}
F_D^{(n)}(a,b_1,\ldots,b_n;c;x_1,\ldots,x_n)\\
	=\sum_{m_1=0}^\infty\cdots \sum_{m_n=0}^\infty
	\frac{(a)_{m_1+\cdots+m_n}(b_1)_{m_1}\cdots(b_n)_{m_n}}{(c)_{m1+\cdots+m_n}m_1!\cdots m_n!}
	\; x_1^{m_1}\cdots x_n^{m_n},
\end{multline}
whenever $\lvert x_1 \rvert,\ldots,\lvert x_n \rvert < 1$ and by analytic continuation
elsewhere. As a matter of notation, $(a)_n=\Gamma(a+n)/\Gamma(a)=a(a+1)\cdots(a+n-1)$ denotes
the Pochhamer symbol (rising factorial).

\subsection{$SU(N)$ without Hypermultiplets}

Using the notation from the previous section, let us choose a canonical homology basis
$B=\{\alpha_i,\beta^i\}_i$ for our curve, such that for any $\gamma\in B$ we have
\begin{equation}
\int_\gamma \lambda_\mathrm{SW}=2\int_{e_i}^{e_j}\lambda_\mathrm{SW},
\end{equation}
for some branch points $e_i$, $e_j$.

Geometrically speaking, the existence of such a basis means that if $\gamma$ encircles
$e_i$ and $e_j$, no other $e$ lies on the straight line connecting those two points, so that
the original contour integral can be converted into twice the integral along that line.

However, for some configurations of the $e$'s no such basis might exist and therefore, unless
stated otherwise, we shall explicitly \emph{assume} its existence, treating it as an
\emph{hypothesis} for what follows.

As stated above, the quantity $a_i$ ($a^i_D$) is obtained by integrating the Seiberg--Witten differential
$\lambda_\mathrm{SW}$ along the cycle $\alpha_i$ ($\beta^i$) for which we indiscriminatingly write $\gamma$.

Thus we have
\begin{equation}\label{result}\begin{split}
a_i\quad \mbox{or} \quad a^i_D &= \int_\gamma\lambda_\mathrm{SW}=
	2\int_{e_i}^{e_j}\lambda_\mathrm{SW}\\
	&=\frac{1}{\pi \mathrm{i}}\int_{e_i}^{e_j}\prod_{k=0}^{N-1}(x-z_k)\prod_{\ell=1}^{2N}
	(x-e_\ell)^{-\frac{1}{2}}\,\mathrm{d}x\\
	&=\frac{1}{\pi \mathrm{i}} (e_j-e_i)\int_0^1\prod_{k=0}^{N-1}\big(e_i-(e_i-e_j)t-z_k\big)
	\prod_{\ell=1}^{2N}\big(e_i-(e_i-e_j)t-e_\ell\big)^{-\frac{1}{2}}\,\mathrm{d}t\\
	&=\frac{1}{\pi \mathrm{i}}(e_j-e_i)\prod_{k=0}^{N-1}(e_i-z_k)(e_i-e_j)^{-\frac{1}{2}}
	(e_j-e_i)^{-\frac{1}{2}}\prod_{\substack{\ell=1\\ \ell\neq i,j}}^{2N}
	(e_i-e_j)^{-\frac{1}{2}}\times\\
	&\quad\times\int_0^1t^{-\frac{1}{2}}(1-t)^{-\frac{1}{2}}\prod_{k=0}^{N-1}
	\left(1-t\frac{e_i-e_j}{e_i-z_\ell}\right)\prod_{\substack{\ell=1\\ \ell\neq i,j}}^{2N}
	\left(1-t\frac{e_i-e_j}{e_i-e_\ell}\right)^{-\frac{1}{2}}\,\mathrm{d}t\\
	&=(e_i-e_j)^\frac{1}{2}\prod_{k=0}^{N-1}(e_i-z_k)
	\prod_{\substack{\ell=1\\ \ell\neq i,j}}^{2N}(e_i-e_\ell)^{-\frac{1}{2}}\times\\
	&\quad\times F_D^{(3N-2)}\Big(\frac{1}{2},
	\underbrace{-1,\ldots,-1}_{N\,\mbox{parameters}},
	\underbrace{\frac{1}{2},\ldots,\frac{1}{2}}_{\substack{2N-2\\\mbox{parameters}}};1;
	\underbrace{\left\{\frac{e_i-e_j}{e_i-z_k}\right\}_k}_{N\,\mbox{variables}},
	\underbrace{\left\{\frac{e_i-e_j}
	{e_i-e_\ell}\right\}_{\ell\neq i,j}}_{2N-2\,\mbox{variables}}\Big)\quad,
\end{split}\end{equation}
where in the last line we have made use of the integral identity \ref{Euler}.

This is the main result of this section.

By a slight change in notation in this formula we can greatly shift emphasis towards a more
`geometric' viewpoint.

To this end we make the following definitions
\begin{align}
	\zeta   & \mbox{$\,:=\,$} e_i-e_j,\\
	\xi_\nu & \mbox{$\,:=\,$} \frac{e_i-e_j}{e_i-z_{\nu-1}},\quad \nu\in\{1,\ldots,N\};\\
	x_\nu   & \mbox{$\,:=\,$} \frac{e_i-e_j}{e_i-e_\nu},\quad \nu\in\{1,\ldots,i-1,i+1,\ldots,2N\}.
\end{align}
Note that $x_i$ is \emph{not defined}.

Furthermore, let
\begin{align}
	b_\xi&\mbox{$\,:=\,$}\underbrace{(-1,\ldots,-1)}_{N\, \mbox{components}},\\
	b_x&\mbox{$\,:=\,$}\underbrace{\big(\frac{1}{2},\ldots,\frac{1}{2}\big)}_{2N-2\,\mbox{components}}.
\end{align}

Then the result \ref{result} reads
\begin{equation}\label{clean}
	a_i \quad \mbox{or} \quad a^i_D=\zeta\;\prod^N_{\nu=1}\xi^{-1}_\nu\,
	\prod^{2N}_{\substack{\nu=1\\ \nu\not=i}}x^{\,\frac{1}{2}}_\nu\;
	F^{(3N-2)}_D\big(\frac{1}{2},b_\xi,b_x;1;\{\xi_\nu\}^N_{\nu=1},
	x_1,\ldots,\widehat{x_j},\ldots,x_{2N}\big),
\end{equation}
where \quad $\widehat{}$ \quad denotes omission. We remark that in total there are $3N-2$
\mbox{`$x$'-arguments} to the $F_D$, $x_i$ being undefined and $x_j$ being omitted.

The promised shift in emphasis takes place if we now consider $a_i$ and $a^i_D$
as functions solely of $\zeta$ and the $\xi$'s and $x$'s, the domain of these
`new' functions being the subset of $\mathbb{C}\times\mathbb{C}^N\times\mathbb{C}^{2N-2}$
which results if we regard $\zeta$ and the $\xi$'s and $x$'s as functions of the $u$'s---the
vacuum expectation values---and let the latter vary freely over the whole of $\mathbb{C}^{N-1}$.

Unfortunately it turns out that the described domain is rather complicated. Nevertheless, we
believe that this viewpoint sheds some additional light on the structure of the moduli space.

The Lauricella function of type $F_D$ has been used in the context of
Seiberg--Witten theory before \cite{Flohr:1998ew,Flohr:2004ug}. However, 
the approach take there made use of conformal invariance of the Seiberg--Witten
periods which leads to a different set of variables. The main problem with
this older approach is that the variables involve crossing ratios of the 
branch points and zeroes of the Seiberg--Witten differential. This makes an
analysis of singular points, as we will perform it in simple examples below,
almost impossible.

\subsection{$SU(N)$ with Hypermultiplets} \label{massive}

As remarked in Section \ref{ETC} the case where massive hypermultiplets are present brings
about some modifications in the Seiberg--Witten differential and the curve on which it is defined. Namely,
if we denote the gauge group by $SU(N_c)$ and if there are present $N_f$ hypermultiplets with masses
$m_r$, the curve-defining equation becomes
\begin{equation}
y^2=A(x)^2-B(x)\mbox{$\,:=\,$}\left(x^{N_c}-\sum_{k=2}^{N_c} u_k x^{N_c-k}\right)^2-
	\Lambda^{2N_c-N_f}\prod_{r=1}^{N_f}(x-m_r)
	=\prod_{i=1}^{2N_c}(x-e_i),
\end{equation}
and the Seiberg--Witten differential then reads
\begin{equation}
\lambda_\mathrm{SW}=\frac{1}{2 \pi \mathrm{i}}
	\frac{\prod_{\ell=0}^{N_c+N_f-1} (x-z_\ell)}
	{\prod_{i=1}^{2N}\sqrt{x-e_i}\prod_{j=1}^{N_f}(x-m_j)}\,\mathrm{d}x,
\end{equation}
where the $z_\ell\;(\ell>0)$ now denote the zeros of $2A'(x)B(x)-A(x)B'(x)$, and again $z_0=0$.

Under the same conditions\footnote{This is not entirely true, because now we must also make
sure, that no $m_r$ lies on the straight line connecting $e_i$ and $e_j$ and we also have to enlarge
the homology basis to include cycles round the poles of the Seiberg--Witten differential as explained in
the introduction. The corresponding scalar modes are suppressed here, since their calculation
reduces to the evaluation of residues.} on our homology
basis as in the previous section we obtain by a calculation absolutely analogous to that of
\ref{result}
\begin{align}
a_i\quad \mbox{or} \quad a^i_D\ &=\ 
          (e_i-e_j)^\frac{1}{2}\prod_{k=0}^{N_c+N_f-1}(e_i-z_k)
	  \prod_{\substack{\ell=1\\ \ell\neq i,j}}^{2N_c}
          (e_i-e_\ell)^{-\frac{1}{2}}
	  \prod_{r=1}^{N_f}(e_i-m_r)^{-1}\times\\
	 &\times F_D^{(3N_c+2N_f-2)}\Big(\frac{1}{2},
          \begin{array}[t]{ccccccc}
	    \underbrace{-1,\ldots,-1}_{N_c+N_f\,\mbox{parameters}}&,&
	    \underbrace{\frac{1}{2},\ldots,\frac{1}{2}}_{
              \substack{2N_c-2\\\mbox{parameters}}}&,&
	    \underbrace{1,\ldots,1}_{
              N_f\,\mbox{parameters}}&;1;&\\
	    \underbrace{\left\{\frac{e_i-e_j}{e_i-z_k}\right\}_k}_{
              N_c+N_f\,\mbox{variables}}&,&
	    \underbrace{\left\{\frac{e_i-e_j}{e_i-e_\ell}\right\}_{
              \ell\neq i,j}}_{
              2N_c-2\,\mbox{variables}}&,&
	    \underbrace{\left\{\frac{e_i-e_j}{e_i-m_r}\right\}_r}_{
              N_f\,\mbox{variables}}& &\Big)\,. 
          \end{array}\nonumber
\end{align}
This is the result for the case with massive hypermultiplets.

Of course one could also clean up the notation in this formula, but since we will not be making
any use of it we shall pass this by.

\section{Applications}
In this section we discuss applications of the general formulas developed earlier. First, as a warm up,
we show how some well known results (namely the asymptotics for the scalar modes $a(u)$ and $a_D(u)$
as $u\rightarrow 0$)
for the $SU(2)$ case can be obtained quite easily. After that, we
study the Argyres--Douglas $\mathbb{Z}_3$-point in the $SU(3)$ case and address in detail the various issues
raised in the introduction.

\subsection{$SU(2)$}\label{su2sect}

Let us calculate the scalar modes when the gauge group is $SU(2)$.
Referring to the curve \ref{curve}, here $N=2$ and there is exactly one modulus $u$.

The $z$'s (cf.\ the line below eq.\ \ref{lambdasw}) are
\begin{equation}
z_0=0,\quad z_1=0
\end{equation}
and the branch points are
\begin{align}
	e_1&=\sqrt{u-\Lambda^2},\\
	e_2&=-\sqrt{u+\Lambda^2},\\
	e_3&=-\sqrt{u-\Lambda^2},\\
	e_4&=\sqrt{u+\Lambda^2}.
\end{align}
Taking this into account, we have the situation depicted in figure \ref{fig1}.

\FIGURE{\includegraphics[width=9cm]{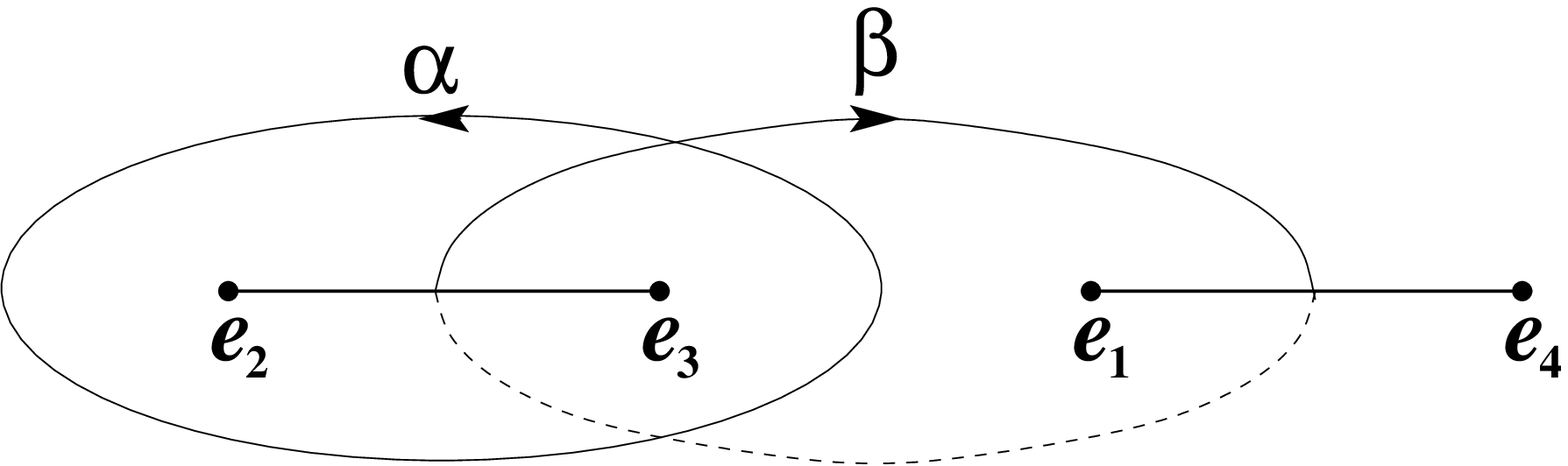}\caption{\label{fig1}Homology basis for $SU(2)$}}

Therefore, using \ref{result} (our homology basis $\{\alpha,\beta\}$ is shown in fig.\ \ref{fig1}),
\begin{equation}\begin{split}
	a(u)&=\int_\alpha \lambda_\mathrm{SW}=-2\int_{e_2}^{e_3} \lambda_\mathrm{SW}\\
	&=-\frac{e_2^2}{(e_2-e_1)^\frac{1}{2}(e_2-e_4)^\frac{1}{2}}\,
	F_D^{(4)}\Big(\ldots;\frac{e_2-e_3}{e_2-0},\frac{e_2-e_3}{e_2-0},\frac{e_2-e_3}{e_2-e_1},
	\frac{e_2-e_3}{e_2-e_4}\Big)\\
	&\sim \frac{1}{\sqrt{8}}\,\sqrt{2u}\,,\quad u \rightarrow \infty,
\end{split}\end{equation}
and
\begin{equation}\begin{split}\label{aD}
	a_D(u)&=\int_\beta \lambda_\mathrm{SW}=-2\int_{e_3}^{e_1} \lambda_\mathrm{SW}\\
	&=-\frac{e_3^2}{(e_3-e_2)^\frac{1}{2}(e_3-e_4)^\frac{1}{2}}\,
	F_D^{(4)}\Big(\ldots;\frac{e_3-e_1}{e_3-0},\frac{e_3-e_1}{e_3-0},\frac{e_3-e_1}{e_3-e_2},
	\frac{e_3-e_1}{e_3-e_4}\Big)\\
	&\sim -\frac{1}{\pi}\frac{e_3^2}{(e_1-e_3)^\frac{1}{2}(e_3-e_4)^\frac{1}{2}}\,
	\log\left(\frac{e_1-e_3}{e_3-e_2}\right),\quad u\rightarrow\infty\\
	&\sim \frac{1}{\sqrt{8}}\,\frac{\mathrm{i}}{\pi}\sqrt{2u}\,\log u,
	\quad u\rightarrow\infty,
\end{split}\end{equation}
where for $a_D(u)$ we have made use of equation \ref{cont} in the appendix.

These results are of course well known.

The rigorous minded reader might want to consider the following paragraph, where we explain in
a little more detail how we applied \ref{cont} to obtain the asymptotic behavior of $a_D(u)$.

\paragraph{Details of previous derivation.}
For the following it is useful to write out in full the $F_D$ in formula \ref{aD}:
\begin{equation}
	F_D^{(4)}\Big(\frac{1}{2},-1,-1,\frac{1}{2},\frac{1}{2};1;\frac{e_3-e_1}{e_3-0},
	\frac{e_3-e_1}{e_3-0},\frac{e_3-e_1}{e_3-e_2},\frac{e_3-e_1}{e_3-e_4}\Big).
\end{equation}
Since the first two arguments are identical and equal to $2$ this reduces to
\begin{equation}
	F_D^{(3)}\Big(\frac{1}{2},-2,\frac{1}{2},\frac{1}{2};1;2,
	\frac{e_3-e_1}{e_3-e_2},\frac{e_3-e_1}{e_3-e_4}\Big),
\end{equation}
as is easily seen from the power series representation \ref{pow}. Using the power series even
though $\frac{e_3-e_1}{e_3-0}\equiv2$ is legal, since the corresponding parameters are negative
integers so that this part of the series terminates after a finite number of terms, giving
something polynomial in the argument.

Following up on the remark `something polynomial' we obtain
\begin{equation}\begin{split}
	F_D^{(3)}\Big(\frac{1}{2},-2,\frac{1}{2},\frac{1}{2};1;\mathbf{2},
	\frac{e_3-e_1}{e_3-e_2},\frac{e_3-e_1}{e_3-e_4}\Big)\\
	=&F_D^{(2)}\Big(\frac{1}{2},\frac{1}{2},\frac{1}{2};1;
	\frac{e_3-e_1}{e_3-e_2},\frac{e_3-e_1}{e_3-e_4}\Big)\\
	-\mathbf{2}&F_D^{(2)}\Big(\frac{1}{2}+1,\frac{1}{2},\frac{1}{2};1+1;
	\frac{e_3-e_1}{e_3-e_2},\frac{e_3-e_1}{e_3-e_4}\Big)\\
	+\frac{3}{8}\mathbf{2}^2 &F_D^{(2)}\Big(\frac{1}{2}+2,\frac{1}{2},\frac{1}{2};1+2;
	\frac{e_3-e_1}{e_3-e_2},\frac{e_3-e_1}{e_3-e_4}\Big).
\end{split}\end{equation}
Now, in the limit $u\rightarrow\infty$, $\lvert\frac{e_3-e_1}{e_3-e_2}\rvert\rightarrow\infty$
and $\lvert\frac{e_3-e_1}{e_3-e_4}\rvert\nearrow1$, the divergence being faster than the
convergence. Therefore we can employ the analytic continuation formula \ref{cont}.
Doing so, one easily verifies that the first asymptotic in \ref{aD} is correct. The expression
in the fourth line is asymptotically equal to the one before and since the relation $\sim$ is
transitive everything is in perfect order.

\subsection{Argyres--Douglas' $\mathbb{Z}_3$-point}\label{application}

In this section we apply our main result \ref{result} to the $\mathbb{Z}_3$-point discovered
by Argyres and Douglas (cf.\ \cite{Argyres:1995jj,Argyres:1995wt}). Also, 
\cite{Gorsky:2000ej,Yung:2001jd,Taylor:2001hg}, and references cited therein, 
have made contributions to this subject. The case with
hypermultiplets was considered in \cite{Ewen:1996uq,Taylor:2002sg}.

Roughly speaking, the Argyres--Douglas point is interesting because it provides
us with an example of a theory in which the BPS spectrum contains a pair of
dual particles, i.e.\ particles where one is electrically and the other 
magnetically charged, which becomes simultaneously massless.
Furthermore, at such points in moduli space, the
theory becomes superconformal which, of course, has important consequences 
\cite{Shapere:1999xr}.

\subsubsection{Vanishing of Scalar Modes}

The $\mathbb{Z}_3$-point is the $e$-configuration in the $SU(3)$ case determined by the choice of
vev's $u=0$, $v=\Lambda^3$ (for completeness, we mention that there exist other choices which also lead
to `$\mathbb{Z}_3$-points').
Our aim is to apply our formula \ref{result} to this case.

Therefore, following Argyres and Douglas, we write
\begin{equation}
u=\delta u ,\quad v=\Lambda^3+\delta v.
\end{equation}

\emph{We shall restrict ourselves to (real) $\delta u<0$ and (real) $\delta v>0$.}
We remark on these hypotheses in Section \ref{discussion} below.

Referring to the curve \ref{curve}, the $e$'s (branch points) are the zeros of
\begin{equation}\label{poly}
\left(x^3-\delta u\,x - (\Lambda^3+\delta v)\right)^2 - \Lambda^6.
\end{equation}
They are easily seen to be separated into two classes: Those which are the zeros of
$p_1(x)=x^3-\delta u\, x-\delta v$ and those which are the zeros of
$p_2(x)=x^3-\delta u\, x -\delta v - 2\Lambda^3$.

Put $P=\mathrm{sgn}(-\delta v/2)\,\sqrt{\lvert -\delta u/3\rvert}$ and
$\beta=\frac{1}{3}\mathrm{arsinh}\frac{-\delta v/2}{P^3}$.

Then, under our assumptions on the $\delta$'s, the zeros of $p_1$ are (see e.g. \cite{bronstein})
\begin{equation}\begin{split}
e_1&=-2P\sinh\beta,\\
e_2&=P(\sinh\beta-\mathrm{i}\sqrt{3}\cosh\beta),\\
e_3&=P(\sinh\beta+\mathrm{i}\sqrt{3}\cosh\beta).
\end{split}\end{equation}
We shall not need explicit formulas for the zeros of $p_2$. It will suffice to know that whenever
the $\delta$'s are small, the zeros of $p_2$ (call them $e_4$,$e_5$,$e_6$) will be near
the third roots of $2\Lambda^3$.

The resulting configuration, along with our choice of homology basis, is visualized in Figure \ref{fig2}.
\FIGURE{\includegraphics[width=11cm]{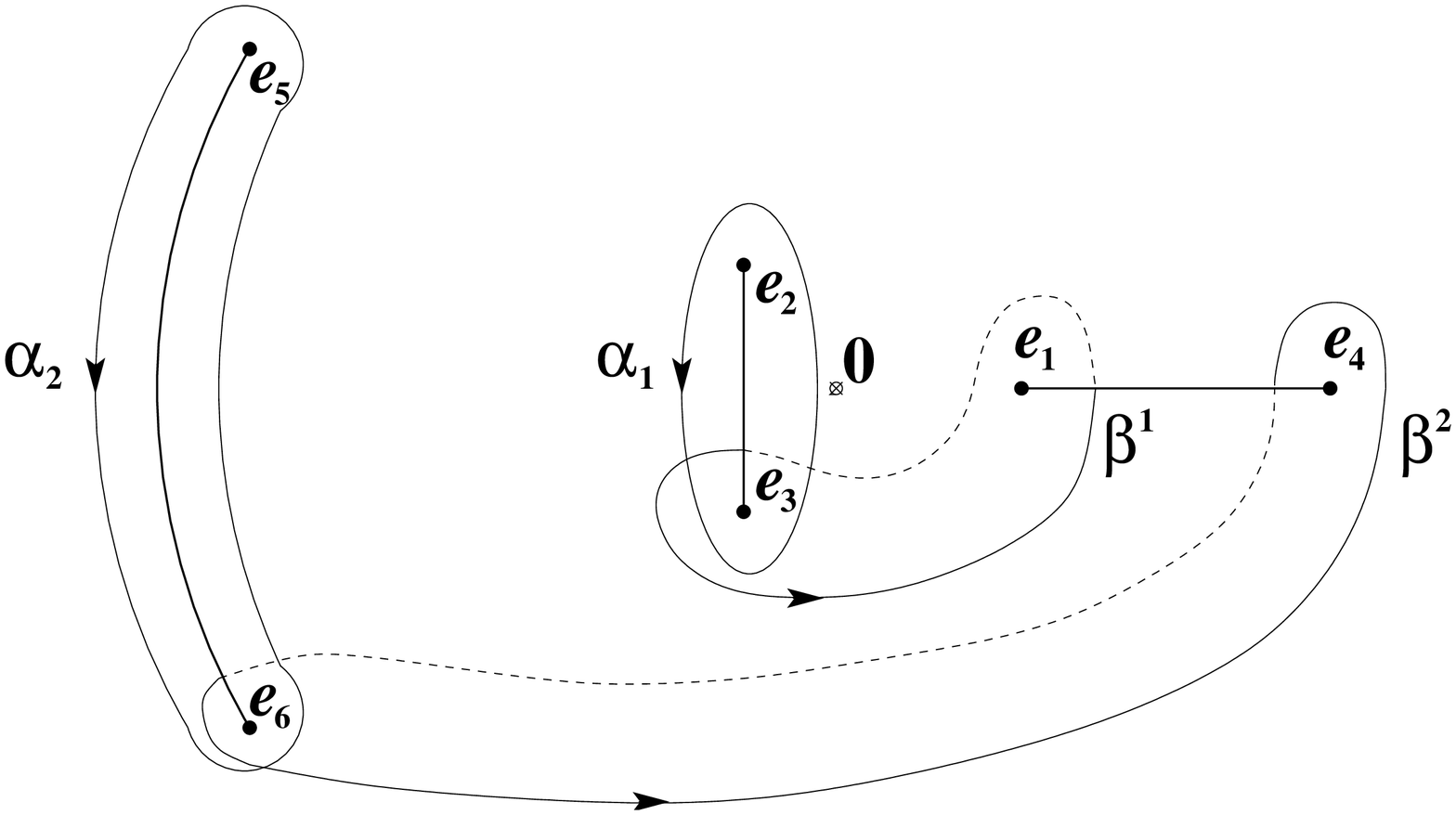}\caption{\label{fig2}Homology basis for $SU(3)$}}
Before writing down our expressions for the $a$'s and $a_D$'s we also need to know that the
$z$'s (recall the definition just below eq. \ref{lambdasw}) now are
\begin{equation}
z_0=0,\quad z_1=\sqrt{\delta u/3},\quad z_2=-\sqrt{\delta u/3}.
\end{equation}

In the study of the $\mathbb{Z}_3$-point one is mostly interested in $a_1$ and
$a^1_D$ because it are these two dual quantities which simultaneously vanish at the
$\mathbb{Z}_3$-point. This, of course, is related to the fact that as the $\delta$'s tend to $0$
the cycles $\alpha_1$ and $\beta^1$ contract to points because $e_1$,$e_2$,$e_3$ tend to $0$.

As functions of $\delta u$ and $\delta v$, i.~e. near the Argyres--Douglas point,
the scalar modes $a_1$, $a^1_D$ are given by (cf. eq. \ref{result})
\begin{multline}\label{su3a}
a_1=2\int_{e_2}^{e_3}\lambda_\mathrm{SW}
	=\frac{e_2(e_2^2-\delta u/3)}
	{(e_2-e_1)^\frac{1}{2}(e_2-e_4)^\frac{1}{2}(e_2-e_5)^\frac{1}{2}(e_2-e_6)^\frac{1}{2}}
	\times\\
	\times F_D^{(7)}\Big(\ldots;\frac{e_2-e_3}{e_2-0},\frac{e_2-e_3}{e_2-\sqrt{\delta u/3}},
	\frac{e_2-e_3}{e_2+\sqrt{\delta u/3}},
	\frac{e_2-e_3}{e_2-e_1},\frac{e_2-e_3}{e_2-e_4},
	\frac{e_2-e_3}{e_2-e_5},\frac{e_2-e_3}{e_2-e_6}\Big),
\end{multline}
and
\begin{multline}\label{su3ad}
a^1_D=2\int_{e_3}^{e_1}\lambda_\mathrm{SW}
	=\frac{e_3(e_3^2-\delta u/3)}
	{(e_3-e_2)^\frac{1}{2}(e_3-e_4)^\frac{1}{2}(e_3-e_5)^\frac{1}{2}(e_3-e_6)^\frac{1}{2}}
	\times\\
	\times F_D^{(7)}\Big(\ldots;\frac{e_3-e_1}{e_3-0},\frac{e_3-e_1}{e_3-\sqrt{\delta u/3}},
	\frac{e_3-e_1}{e_3+\sqrt{\delta u/3}},
	\frac{e_3-e_1}{e_3-e_2},\frac{e_3-e_1}{e_3-e_4},
	\frac{e_3-e_1}{e_3-e_5},\frac{e_3-e_1}{e_3-e_6}\Big).
\end{multline}

The next thing we shall do is to examine the behavior of $a_1$ and $a^1_D$ as the $\delta$'s both
tend to $0$. There are several meanings one can attach to the phrase `tend to $0$.'
One possible meaning is the usual one from the calculus of several (complex) variables, where
both variables are considered as \emph{independent} of each other.
Another possible meaning, and this is often done in physics, is to consider both variables
as \emph{dependent}, say $\delta v=-\delta u$ (recall our assumptions about the $\delta$'s).
We will follow the latter course (cf. our general remarks in Section \ref{discussion} below).
This ensures that both variables will be of the same order of smallness during the limiting
process.
\paragraph{We first consider $a_1$:}
As $\delta u \rightarrow 0$, $\delta u<0$, the arguments to the $F_D$ in eq. \ref{su3a} behave as follows.
\begin{equation}\begin{split}\label{limits}
\frac{e_2-e_3}{e_2-0}&\rightarrow\frac{1}{2}(3-\mathrm{i}\sqrt{3}),\\
\frac{e_2-e_3}{e_2-\sqrt{\delta u/3}}&\rightarrow\frac{1}{2}(3-\mathrm{i}\sqrt{3}),\\
\frac{e_2-e_3}{e_2+\sqrt{\delta u/3}}&\rightarrow\frac{1}{2}(3-\mathrm{i}\sqrt{3}),\\
\frac{e_2-e_3}{e_2-e_1}&\rightarrow\frac{1}{2}(1-\mathrm{i}\sqrt{3}),\\
\frac{e_2-e_3}{e_2-e_4}&\rightarrow 0,\\
\frac{e_2-e_3}{e_2-e_5}&\rightarrow 0,\\
\frac{e_2-e_3}{e_2-e_6}&\rightarrow 0.
\end{split}\end{equation}
Furthermore, \emph{the prefactor of the $F_D$ in eq. \ref{su3a} tends to $0$.} These limits
were obtained using {\sc Mathematica}.
We will now prove that the value of the $F_D$ at the limits just written down is a
complex number (i.~e., that it is defined for that particular constellation of arguments).

First observe that
\begin{equation}
F_D^{(7)}(\ldots,0,0,0)=F_D^{(4)}(\ldots),
\end{equation}
and
\begin{multline}
F_D^{(4)}\Big(\frac{1}{2},-1,-1,-1,\frac{1}{2};1;\frac{1}{2}(3-\mathrm{i}\sqrt{3}),
\frac{1}{2}(3-\mathrm{i}\sqrt{3}),\frac{1}{2}(3-\mathrm{i}\sqrt{3}),\frac{1}{2}(1-\mathrm{i}\sqrt{3})\Big)\\
=F_D^{(2)}\Big(\frac{1}{2},-3,\frac{1}{2};1;\frac{1}{2}(3-\mathrm{i}\sqrt{3}),\frac{1}{2}(1-\mathrm{i}\sqrt{3})\Big).
\end{multline}
Since $-3$ is a negative integer this last $F_D$ reduces to a polynomial in $\frac{1}{2}(3-\mathrm{i}\sqrt{3})$ (see
Appendix \ref{appendix}), the
coefficients being rational numbers multiplied by expressions of the form\linebreak
${}_2F_1\left(\frac{1}{2}+n,\frac{1}{2};1+n;\frac{1}{2}(1-\mathrm{i}\sqrt{3})\right)$, where $n$ is some
nonnegative integer (${}_2F_1$ is the Gaussian hypergeometric function). Since for our constellation of
parameters the only singularity of
${}_2F_1$ on the unit circle is at $1$, these `expressions' reduce to (finite) complex numbers.
Thus, as promised, the $F_D$ in eq. \ref{su3a} converges at the limits of its arguments.

\emph{Therefore (recall that the prefactor tends to $0$), $a_1$ tends to $0$, as $\delta u \rightarrow0$,
$\delta u<0$.}
\paragraph{We now consider $a^1_D$:}
Referring to eq. \ref{su3ad} one finds that the arguments of the $F_D$ in that equation, taken in
the same order as those of the $F_D$ in eq. \ref{su3a}, tend to the same limits as indicated in
eq. \ref{limits}. Following exactly the same arguments as for $a_1$ we find that the $F_D$ in the
expression for $a^1_D$ tends to the same (finite) complex number as the one in eq. \ref{su3a}.
Also, the prefactor in eq. \ref{su3ad} tends to $0$ as $\delta u\rightarrow0$, $\delta u<0$.

\emph{Therefore, $a^1_D$ tends to $0$ as $\delta u\rightarrow0$, $\delta u<0$.}

This vanishing of $a_1$ and $a^1_D$ reproduces the result stated by Argyres and Douglas \cite{Argyres:1995jj}
in the special limit (recall our restrictions on the $\delta$'s) we have considered.

\subsubsection{General Discussion}\label{discussion}
The reader might ask why in the previous section we imposed so peculiar restrictions
on $\delta u$ and $\delta v$. In the present section, we wish to address precisely this question.

Partly the answer is simple: We imposed these conditions so that we were able to consider the limit
of the scalar modes at the Argyres--Douglas point in a straight-forward manner.
For instance, we know for sure that it is possible to consider the case $\delta u>0$,
$\delta u=\delta v$, $\delta u \rightarrow 0$,\footnote{Again, $a_1$ and $a^1_D$ vanish.} and the analysis of
this is exactly the same as in the case we
have presented in detail (even the arguments of the $F_D$'s tend to the same limits). However, there is a
complication involved. Now one has to look at the discriminant of $p_1$; if it is $0$,
a degeneration of the `inner' $e$'s (cf. Fig. \ref{fig2}) takes place ($e_2=e_3$, moreover,
$e_2=-\sqrt{-\delta u/3}$), and if it is negative, the homology basis of Fig.~\ref{fig2}
is inadequate for our formulas because then all `inner' $e$'s are real. One does away with the trouble
of the discriminant in this case by setting $\delta u=\delta v$ (this implies positivity of the discriminant).
The moral is that it is possible to impose different restrictions on the $\delta$'s than we have done in this paper.

In the general case of arbitrary complex $\delta u, \delta v$ with an arbitrary approach to $0$ (as in
standard calculus) the complications pile up: Firstly, there is the technical problem of keeping track of
third roots of complex numbers. This is very difficult, since the expressions involved in the usual Cardano's formula
are intrinsically discontinuous due to branch cuts; also, multivaluedness becomes a burden.
Secondly, there is the problem that one cannot choose a
homology basis once and for all. For instance, if $\delta v>0$, $\delta u<0$ the configuration of the `inner' $e$'s
looks like that of those in Fig.~\ref{fig2}, only reflected through the origin. Of course this necessitates the
use of a different homology basis than the one shown in Figure \ref{fig2}. If the $\delta$'s can arbitrarily
approach $0$, then it is clear that one could start with a homology basis like the one in Fig. \ref{fig2}
and end up with the need to chose a different one. Our formulas are not suited for such a case.

The question arises: Do $a_1$,$a^1_D$ vanish at the Argyres--Douglas point regardless of exactly how
the $\delta$'s tend to $0$? If they should happen to be continuous functions the answer is affirmative: Yes!
However, we do not deem their continuity as self-evident, since the $e$'s depend on the $\delta$'s in a
rather complicated fashion. On some Riemann-surface $a_1$,$a^1_D$ most likely \emph{are} continuous. But that
seems to be something different.

It is tautological to say that if $a_1(\delta u, \delta v)$,$a^1_D(\delta u, \delta v)$ should happen
to be discontinuous at $(0,0)$, then there would exist some approach of the $\delta$'s to $0$ which would
not yield a vanishing limit.

In light of the particular properties of the Lauricella functions under
analytic continuation, we point out the following: The periods $a_1$ and
$a^1_D$ are given in terms of Lauricella functions $F_D^{(7)}$ of seven
arguments. As can be inferred from eqs.\ \ref{su3a} and \ref{su3ad}, the
arguments depend on the branch points. In our conventions, the three branch
points $e_1$, $e_2$ and $e_3$ are all located in a small neighbourhood around
$0$, if we approach the Argyres--Douglas point. However, generically, they
are farther away from $0$ than the two other zeroes of the Seiberg--Witten
differential $z_{1,2}=\pm\sqrt{u/3}$. It is an easy task to convince oneself,
that appropriate choices of $\delta u$ and $\delta v$ can invert this situation,
such that some of the `small' branch points are closer to zero than the
$z$'s. Of course, then the situation is possible where a branch point coincides
with one of the $z$'s before the point $0$ is approached. 
Also, it can be arranged that two of the three `small' branch
points coincide first, before the three finally approach zero.
All these situations fall in the class of singular sub-manifolds.

Let us look specifically at the fourth argument of the periods. In the
notation introduced in section 2.2 it reads
$x_1=\frac{e_2-e_3}{e_2-e_1}$ for $a_1(\delta u,\delta v)$ or 
$x_2=-\frac{e_3-e_1}{e_2-e_3}$ for $a_D^1(\delta u,\delta v)$, respectively.
Clearly, depending on how the three `small' branch points approach each
other and $0$, the moduli of the two arguments are related as 
$|x_1|=1/|x_2|$ and hence are located around one of the following pairs of
singular points: $(1,1)$, $(0,\infty)$, or $(\infty,0)$.
Since this fourth argument is associated with a parameter $b_x$, which is not
a negative integer, the corresponding Lauricella function is not simply
polynomial in this argument. Thus, in all four cases will we need at
least one analytic continuation, either of $a_1$ or of $a_D^1$. Due to
the specific values of the paramters, each of these analytic continuations
will involve a logarithmic divergence of the involved Lauricella function,
killed off only by the prefactor. 

Next, we look at the last three arguments. Generically, all three approach
$0$ from three different directions. However, in particular if $\Lambda$ is
allowed to be small and complex, the `large' branch points might run into 
each other first, e.g.\ on the sub-manifold characterized by
$\{\delta u,\delta v,\Lambda,w:\delta u=3w^2,\delta v+2\Lambda^3=-2w^3\}$. 
This is then exactly the difficult situation that the singular
point $(x_{1,2},0,0,0)$ is a common point of various sub-manifolds $x_k=x_l$ for
$k,l\in\{4,5,6\}$. Similarly, the first three arguments close in on one
common singular point, but this is not really a difficulty, since the
corresponding parameters render the Lauricella functions polynomial in their
first three variables. Hence, analytic continuation in one of the first three 
variables is only necessary in the true limit of the Argyres--Douglas point
itself, where these variables all tend to infinity. For this, one should note
that the zeroes of $\lambda_{{\rm SW}}$ are generically closer to $0$,
corresponding to the Argyres--Douglas point, than the `small' branch points. 

Therefore, depending on the choice of the $\delta$'s, we may move on or across 
singular sub-manifolds which will make it necessary to perform the appropriate
analytic continuations. In particular, if we have to analytically continue in
more than one variable, say $x_5$ and $x_6$, great care has to be taken that 
for an analytic continuation around, say, the point $(1,1)$ we need {\em two}
different continuations, namely for the patches $|x_5|<|x_6|$ and 
$|x_5|>|x_6|$, respectively. These can be found along the lines set out in
the appendix, but they are more complicated than the examples provided there.
The discriminant of the curve $y^2$ tells us
where branch points coincide. A special locus of that kind is
$\{\delta u,\delta v:27(\delta v)^2=4(\delta u)^3\}$, where two of the three 
`small' branch points become
equal, say $e_2$ and $e_3$. Then $a_1(\delta u,\delta v)$ is well defined
around the singular point $(0,0,0,0)$ for its last four variables, while
$a_D^1(\delta u,\delta v)$ has to be analytically continued around 
$(\infty,0,0,0)$ according to \ref{a:00i}. The situation gets more involved,
if we allow $\Lambda$ to vary arbitrarily or if we include massive 
hypermultiplets. Then, almost any singular sub-manifold can be reached, and
the distinction between `small' and `large' branch points may cease
to exist, leading to higher-dimensional singular sub-manifolds.

Lastly, we would like to emphasize that eqs. \ref{su3a} and \ref{su3ad} are, of course, valid without any particular
restrictions on the $\delta$'s other than that all relevant expression be defined.

\section{Conclusions and Outlook}

We have derived general formulas for the scalar modes in Seiberg--Witten theory for gauge group
$SU(N)$ in the absence as well as in the presence of massive hypermultiplets. These formulas involve
the Lauricella function $F_D^{(n)}$, a multiple hypergeometric function.
The advantage of our approach is threefold. Firstly, our formulas can be
worked out directly, without the need to first find the correct system of
Picard--Fuchs differential equations and its solutions. Secondly, our
formulas work equally well for the case with (massive) hypermultiplets,
putting them on equal footing with the ordinary case. Finally, one can
immediately extract a well-defined power series expansion for any BPS
period from our results, as long as the corresponding homology cycle is
the smallest one.\footnote{This is a sufficient condition. Of course, easily
computable power series expansions exist for many more cases.} Thus, in any
given patch of moduli space, the lightest BPS state for this patch can be
obtained in a straight-forward manner.

Subsequently, we have applied the formula for the case without hypermultiplets to the Argyres--Douglas
point for gauge group $SU(3)$.
Specifically, we have considered the limits of the scalar modes corresponding to the
cycles $\alpha_1$,$\beta^1$ as the vacuum expectation values $u$,$v$ tend---in a special way---to those values which
define the Argyres--Douglas point. In accord with the result of Argyres and Douglas we have found that
the limit of both these modes is indeed zero.

As already pointed out, this raises the question of what happens when $u$,$v$ \emph{arbitrarily} tend to the
prescribed values.
We cannot tell for sure. To begin with, we think it is conceivable that there exists some mode of approach
which yields a limit other than $0$. This possibility is intimately connected with---indeed,
is equivalent to---the question whether the scalar modes are continuous functions of the vacuum expectation values.
With the existence of phase transitions in Seiberg--Witten theory in mind, we believe that it is necessary to
prove the continuity of the scalar modes rather than just assume it.

Even if the scalar modes should turn out to be continuous it is quite puzzling to observe that in the special
case we have considered the Lauricella functions involved in our formulas tend to the same values, so that
the vanishing of the scalar modes is determined by the vanishing of the prefactors. What is puzzling about this
is that it looks as if $a_1^{}$,$a^1_D$ become linearly dependent at the Argyres--Douglas point. In light of
the theory of Lauricella functions we conjecture that the set of BPS periods
alone is not exhaustive to define the scalar modes everywhere in moduli space.
Furthermore, if indeed $a_1^{}$ and $a^1_D$ become linearly dependent, then we 
might happen to be sitting on the curve of marginal stability, meaning that not both of 
the corresponding BPS modes actually exist.

As discussed, it is a known problem in the theory of generalized hypergeometric
functions of several variables that analytic continuation around singular
points becomes extremely difficult. The reason is the appearance of singular
sub-manifolds in case more than one of the variables approaches a singluar
point, or more than two variables approach each other. It is interesting to
note the following: In case of gauge group $SU(N)$ without the presence of
hypermultiplets, the hyperelliptic curve can be written as $y^2 = A(x)-
\Lambda^{2N}$. The Seiberg--Witten differential then is proportional to
$\frac{1}{y}A'(x)x{\rm d}x$. Thus, the zeroes of the Seiberg--Witten
differential are given precisely by those sub-manifolds, where $y^2=A(x)^2-
\Lambda^{2N}$ has a double zero, i.e.\ where we run into difficulties with 
ordinary analytic continuation techniques.\footnote{
The situation is more complicated in the presence of massive hypermultiplets.}
The special zero $x=0$ naturally
arises in the semi-classical regime $\Lambda\rightarrow 0$. The reader
should keep in mind that the zero of $\lambda_{{\rm SW}}$ which we 
studied in our treatment of the Argyres--Douglas point is actually the
zero $x=0$ which always (not only in the semi-classical regime) arises
for $u_N=\Lambda^N$ in $A(x)=x^N-\sum_{k=2}^Nu_kx^{N-k}$, for which $y^2=0$.
Thus, this is actually such a double zero of $y^2$.

In our approach, zeroes of the Seiberg--Witten differential appear on equal
footing with the branch points of $y$ and the poles, in case massive 
hypermultiplets are included. However, when explicit computations of the
BPS periods are needed, the zeroes can be used to decompose the
Lauricella functions of $n$ arguments into linear combinations of such with
fewer arguments, since the dependency on a zero $x=z_\ell$ is always only
polynomial. On the other hand, it has been pointed out 
\cite{geodesics1,geodesics2,geodesics3} that the zeroes of the Seiberg--Witten 
differential have a physical meaning. Our analysis suggests that the
zeroes of the Seiberg--Witten differential are precisely related to the
sub-manifolds which delimit the hypercones in which analytic continuations
can be uniformly defined. Schulze and Warner \cite{geodesics2}
introduce geodesic integration paths which start or end in zeroes of the
Seiberg--Witten differential. In principle, we can extend our analysis by
{\em formally\/} enlarging our homology to include the zeroes of the
Seiberg--Witten differential as further branch points of a thus implicitly
defined Riemann surface. This is usually achieved by deforming the
associated paramters of the Lauricella functions slightly away from negative 
integers and taking appropriate limits afterwards. However, there is presumably
an alternative point of view. As we have shown, analytic continuations of the
Lauricella functions of type $D$ cannot be obtained everywhere within the
set of these functions alone. Enlarging the set of functions confronts one 
with new types of generalized hypergeometric functions which do not possess
integral representations in terms of line integrals or Pochhammer double loop
integrals, where the loop encloses two and only two singular points. 
It is conjectured in the literature (see e.g.\ \cite{Exton:yx}) that one
needs at least the set of all double circuits and all three-foil loops, which
are selfintersecting loops which enclose two or three disjoint subsets of 
singular points, respectively.

We have used precisely the form of the Seiberg--Witten differential which
arises in the context of string theory as a local form of a metric on the
auxiliary Riemann surface $\Sigma$. Although we have followed the traditional 
approch where BPS states are defined in terms of periods of homology elements,
we have found that the set of homology cycles is neither sufficient, nor everywhere
suitable, to describe the BPS spectrum. We do not know how to interpret 
analytic continuations of the scalar modes, which do not have nice and simple
integral representations, but it strikes us as remarkable that regions with
given analytic continuations have the form of hypercones delimited by
sub-manifolds, which are parametrized by the zeroes of the Seiberg--Witten
differential. The theory of Lauricella functions seems at present not well developed
enough to fully understand how geodesic integrals between zeroes of the
Seiberg--Witten differential can be expressed in our approach, i.e.\ in terms
of Lauricella functions and in particular their analytic continuations valid
on the singular sub-manifolds. It is worth noting in this context that
Lauricella functions do have the special property that line integrals of the
Seiberg--Witten differential to an arbitrary point in the complex plane can
easily be written down, namely
\begin{align}
  \int_0^z{\rm d}u\,&u^{a-1}(1-u)^{c-a-1}\prod_{i=1}^n(1-ux_n)^{-b_n} 
  \mbox{$\,=:\,$}
  \frac{\Gamma(a)\Gamma(c-a)}{\Gamma(c)}{\,}^{{\rm inc}\!}_{\ z}F_D^{(n)}
  (a,b_1,\ldots,b_n;c;x_1,\ldots,x_n)
  \nonumber\\
  &=
  \frac{z^a}{a}F_D^{(n+1)}(a,b_1,\ldots,b_n,1+a-c;a+1;x_1,\ldots,x_n,z)
  \,,
\end{align}
which is known also as the incomplete Lauricella function.
If $|z|>1$, this expression has to be replaced by the appropriate analytic 
continuation. 

Therefore, linking our analysis to the study of geodesic horizons would
be a most interesting direction for future research. More work is needed
in order to fully understand the analytic properties of the Lauricella
system $F_D^{(n)}$. Furthermore, we have a price to pay for being able to
compute periods of the Seiberg--Witten form explicitly: We have to change
the variables from the vev's of the scalar field to quotients of differences
of branch points. This is a highly non-linear and multi-valued map.
However, we would like to point out that the alternative, where one works
with the vev's $u_k$, means to find the system of Picard--Fuchs equations and
to solve it. This has been achieved in the case of gauge group $SU(3)$,
\cite{Klemm:1994qs,Klemm:1995wp}, but the results are given in terms
of the Appell function $F_4$ of two arguments (which are relatively simple
functions of $u_1$ and $u_2$), which does not admit 
a nice integral representation. Furthermore, the problem of finding analytic
continuations is similarly complicated. There are regions in moduli space
where simple monodromy transformations will yield the desired results,
but a complete set of analytic continuations cannot be obtained in this
way. Of course, the reason is again the existence of higher-dimensional
singular sub-manifolds, here in $\mathbb{C}^2$ spanned by $u_1$ and $u_2$.
In addition, the Appell function $F_4$ and its generalization to more than
two variables has an even worse structure under analytic continuation than
the Lauricella $F_D^{(n)}$ system. Therefore, we believe that our approach offers
several advantages when we wish to investigate Seiberg--Witten theory in
all of its moduli space.

First of all, the Lauricella system $F^{(n)}_D$ together with the related 
functions $D^{(n)}_{p,q}$ are the simplest of all the generalized
hypergeometric systems of several variables. It might be tedious, but it is
possible to compute analytic continuations for them such that one can
access all of moduli space, not only the semi-classical regime. Moreover,
the formulas are explicit and thus allow direct computations of the
scalar modes.

Secondly, the geometrical setup behind our approach makes the appearance
of singular sub-manifolds very transparent. Since we realize the 
Seiberg--Witten curve as a ramified covering of the complex plane, the branch
points, zeroes and poles of the Seiberg--Witten form have a direct meaning.
However, we lose the direct relation to the vev's of the scalar field.
It depends on what one wishes to consider as `fundamental.' From the point
of view of effective field theory, the vev's are the physically given parameters,
and the Riemann surface is an auxiliary construction. From the vista of
string theory, the Riemann surface has a certain reality. The Seiberg--Witten
form is then uniquely fixed in the local form we used and which serves as a
metric on $\Sigma$.

What remains open is the question how BPS states should really be
characterized. The traditional approach, which also was our starting point,
is via homology cylces of the Riemann surface $\Sigma$. Vanishing cycles
then indicate BPS states which become massless. Other approaches emphasize
the role of geodesic integration paths and zeroes of the Seiberg--Witten
form. We believe that our approach is well suited to both points of view.
It even suggests that just which 
characterization is more appropriate depends on the region of moduli space 
under consideration. A more detailed analysis of this question
is very important and will be investigated in a forthcoming work.

\acknowledgments
N.A.\ thanks Christian S\"amann for useful discussions.
M.F.\ thanks Albrecht Klemm and Werner Nahm for useful discussions.
The work of M.F.\ is supported by the European Union network 
HPRN-CT-2002-00325 (EUCLID) and in part by the string theory 
network (SPP no.\ 1096), Fl 259/2-2, of the Deutsche Forschungsgemeinschaft.

\appendix
\section{Lauricella {\boldmath $F_D^{(n)}$}}\label{appendix}
The purpose of this appendix is to collect various information pertaining to Lauricella
$F_D^{(n)}$. Some of the formulas have already made their appearance in one or the other
section of the main text.
The major reference for Lauricella $F_D^{(n)}$ and other `multiple hypergeometric 
functions' is \cite{Exton:yx}, from which we will cite freely.

\subsection{The Definition}
Lauricella $F_D^{(n)}$ is a function of $n$ complex variables and $n+2$ parameters, defined
by the power series
\begin{multline}\label{power}
F_D^{(n)}(a,b_1,\ldots,b_n;c;x_1,\ldots,x_n)\\
	=\sum_{m_1=0}^\infty\cdots \sum_{m_n=0}^\infty
	\frac{(a)_{m_1+\cdots+m_n}(b_1)_{m_1}\cdots(b_n)_{m_n}}{(c)_{m1+\cdots+m_n}m_1!\cdots m_n!}
	\; x_1^{m_1}\cdots x_n^{m_n},
\end{multline}
whenever $\lvert x_1 \rvert,\ldots,\lvert x_n \rvert < 1$ and by analytic continuation
elsewhere. The symbol $(a)_n=\Gamma(a+n)/\Gamma(a)$ is the so-called Pochhammer symbol.

The function $F_D^{(n)}$ has the integral representation
\begin{equation}
\int_0^1 t^{a-1}(1-t)^{c-a-1}\prod_{i=1}^n(1-tx_i)^{-b_i}\, \mathrm{d}t=
	\frac{\Gamma(a)\Gamma(c-a)}{\Gamma(c)}\,F_D^{(n)}(a,b_1,\ldots,b_n;c;x_1,\ldots,x_n),
\end{equation}
if $\mathrm{Re}(a)$ and $\mathrm{Re}(c-a)$ are positive. This is proved using the binomial theorem. See
\cite{Flohr:2004ug}, Appendix~B for a detailed derivation.

A number of facts can be read off the power series representation \ref{power}:
\begin{enumerate}
\item If one of the variables of $F_D^{(n)}$, say $x_i$, is equal to $0$, then
the $F_D^{(n)}$ reduces to a $F_D^{(n-1)}$:
\begin{multline}
F_D^{(n)}(a,b_1,\ldots,b_i,\ldots,b_n;c;x_1,\ldots,x_{i-1},0,x_{i+1},\ldots,x_n)\\
=F_D^{(n-1)}(a,b_1,\ldots,\widehat{b_i},\ldots,b_n;c;x_1,\ldots,x_{i-1},x_{i+1},\ldots,x_n),
\end{multline}
where \quad $\widehat{}$ \quad denotes omission.

\item If two variables have equal values, say $x_i=x_j$, a similar reduction takes place:
\begin{multline}
F_D^{(n)}(a,b_1,\ldots,b_i,\ldots,b_j,\ldots,b_n;c;x_1,\ldots,x_i,\ldots,x_j,\ldots,x_n)\\
=F_D^{(n-1)}(a,b_1,\ldots,b_{i-1},b_i+b_j,b_{i+1},\ldots,\widehat{b_j},\ldots,b_n;c;x_1,\ldots,x_i,
\ldots,\widehat{x_j},\ldots,x_n).
\end{multline}

\item If $b_i$ is a negative integer, then the part of the series corresponding to $x_i$ terminates after a finite
number of terms (because $(b_i)_n=0$ for $n>\lvert b_i \rvert$) and thus reduces to a polynomial in $x_i$.
In this case, the modulus of $x_i$ is immaterial for the validity of \ref{power}.
\end{enumerate}

If $\lvert x_i \rvert\geq 1$, for some $i\in\{1,\ldots,n\}$, then the power series representation \ref{power}
is not valid anymore. Rather, one must have recourse to an analytic continuation of $F_D^{(n)}$.
This is effected by writing
\begin{multline}\label{contprep}
F_D^{(n)}(a,b_1,\ldots,b_n;c;x_1,\ldots,x_n)\\
=\sum_{m_1=0}^\infty\cdots\sum_{m_{i-1}=0}^\infty\sum_{m_{i+1}=0}^\infty\cdots \sum_{m_n=0}^\infty
\frac{(a)_{m_1+\cdots+m_{i-1}+m_{i+1}+\cdots+m_n}\prod_{\ell\neq i}(b_\ell)_{m_\ell}}
	{(c)_{m_1+\cdots+m_{i-1}+m_{i+1}+\cdots+m_n}\prod_{\ell\neq i}m_\ell!}\prod_{\ell\neq i}x_\ell^{m_\ell}\times\\
	\times{}_2F_1(a+\sum_{\ell\neq i}m_\ell,b_i;c+\sum_{\ell \neq i}m_\ell;x_i),
\end{multline}
and employing a suitable continuation formula for the Gaussian hypergeometric function ${}_2F_1$.
Erd\'elyi~et~al. \cite{Erdelyi} as well as Becken and Schmelcher \cite{Becken} give several such continuation formulas.
Especially the latter reference appears to be quite exhaustive. Also, see the next section \ref{continuation}.

Nevertheless, sometimes one can get along without any explicit continuation formula, just as we have done in Section
\ref{application}.

\subsection{Analytic continuation of Lauricella functions}\label{continuation}

The Seiberg--Witten periods are analytic functions `everywhere' in the
moduli space, i.e.\ for generic values of either the vacuum expectation
values $u_k$ or the branch points $e_\ell$. However, it is clear that this is
not necessarily the case in the singular regions where one or more branch
points $e_\ell$ become identical. In fact, a typical feature of a dual
pair $(a,a_D)$ of Seiberg--Witten periods corresponding to a dual pair of homology cycles
$(\alpha,\beta)$ is that with a vanishing cycle $\alpha\rightarrow 0$, only
$a(u)$ becomes small, while the dual period $a_D(u)$ diverges logarithmically.

Precisely this is it what makes the Argyres--Douglas point so interesting: a pair
of dual periods \emph{simultaneously} become small, i.e.\ two particles 
dual to each other under some sort of electromagnetic duality,
simultaneously become massless. As a consequence of this the theory is
conformally invariant in this regime.

There is another interesting fact about such points in moduli space, where
intersecting homology cycles vanish at the same time. If one uses the
branch points as the natural coordinates to parametrize the theory, it
is known from the theory of generalized hypergeometric functions
that a complete set of analytic continuations cannot be given
entirely in terms of the same class of functions one starts out with (cf.\  \cite{Exton:yx}).
This is to be contrasted with the well-known result that the ordinary
Gaussian hypergeometric function admits analytic continuations everywhere in
the complex plane, which again can be expressed as linear combinations of
Gaussian hypergeometric functions (cf., e.g.\ \cite{Becken}). This can be used, for example, to
find analytic continuation formulas for the Lauricella $F_D^{(n)}$ function, as
long as we only need to continue one of its arguments outside the unit
circle of convergence. As soon as we wish to have more than one argument
outside the unit circle, things become complicated. In case of the
Lauricella $F_D^{(n)}$ function, one is confronted with the following problem:

Within its region of convergence, the Lauricella $F_D^{(n)}$ series can be
represented in the form of an Euler--type integral, i.e.\ an integral along
a simple loop, which we choose to be one of the homology cycles. More
generally, Pochhammer double loops might be admitted as well. The point
is, that only two of the singular points of the differential are
enclosed by the loop. The thus defined functions possess analytic 
continuations which, for generic values of the parameters, can again be
given in terms of multi-variable power series (perhaps multiplied by a common 
fractional power). However, such analytic continuations are only valid
outside the unit ball within cones delimited by the singular hyperplanes,
given by coinciding singular points. Thus, one needs a considerably
larger set of analytic continuations than in the single-variable case.
Morevoer, not all of these analytic continuations can be represented by
Euler--type integrals. This raises the physically relevant question what
the meaning of the Seiberg--Witten periods then is. As long as they can be
understood as contour integrals along homology basis elements, they represent
the mass of particles with a charge fixed by the corresponding homology element. But what
would the meaning be, if no such simple contour existed? Exton \cite{Exton:yx}
mentions that at least so-called three-foil loops are necessary to be able
to represent a full set of analytic continuations in terms of integrals.
Three-foil loops are three times self-intersecting loops which enclose three
different sets of singular points. 

The linearity of the integral guarantees that, in principle, we could eliminate
the internal dependence on the zeros of the Seiberg--Witten differential by
decomposing the integrand. The resulting Lauricella functions of fewer
arguments depend solely on the branch points via the quotients of
differences of them, as introduced earlier. However, generic small variations
of a subset of branch points which are in a small neighborhood around a
physically interesting singular point such as the Argyres--Douglas point,
have to be treated carefully. On the one hand, we might need several analytic
continuations, because the small variation crosses a boundary of the cone of
convergence, which coincides with the situation that the initial choice of
a homology basis ceases to be valid. On the other hand, we even might run into
a region where no Euler--type integral representation exists.

In order to study the analytic continuations of the Lauricella functions of
type $F_D$, one needs a further class of related functions, defined by
the expansions
\begin{multline}\label{Dpower}
D^{(n)}_{p,q}(a,b_1,\ldots,b_n;c,c';x_1,\ldots,x_n)\\
	=\sum_{m_1=0}^\infty\cdots \sum_{m_n=0}^\infty
	\frac{(a)_{m_{p+1}+\cdots+m_n-m1-\cdots-m_p}(b_1)_{m_1}\cdots(b_n)_{m_n}}{(c)_{m_{q+1}+\cdots+m_n-m_1-\cdots-m_p}c'_{m_{p+1}+\cdots+m_q}m_1!\cdots m_n!}
	\; x_1^{m_1}\cdots x_n^{m_n},
\end{multline}
where $0\leq p\leq q\leq n$. It is important to note that these 
functions, which appear in the analytic continuations of the Lauricella
$F_D^{(n)}$ functions, do not possess Euler--type integral representations.
The simplest known integral representation is in fact a Pochhammer double loop
integral involving a Lauricella function in its kernel, namely
\begin{multline}\label{Dint}
\frac{(2\pi{\rm i})^2}{\Gamma(a)\Gamma(a')\Gamma(2-a-a')}
D^{(n)}_{p,q}(a+a'-1,b1,\ldots,b_n;c,c';x_1,\ldots,x_n)\\
= \int{\rm d}u\,(-u)^{-a'}(u-1)^{-a}F_D^{q-p}(a',b_{p+1},\ldots,b_q;c';
  \frac{x_{p+1}}{u},\ldots,\frac{x_q}{u})\times\\ 
\times D^{(p)}_{n-q+p,n-q+p}(a,b_{q+1},\ldots,b_n,b_1,\ldots,b_p;c,c;
  \frac{x_{q+1}}{1-u},\ldots,\frac{x_n}{1-u},\frac{x_1}{1-u},\ldots,
  \frac{x_p}{1-u}),
\end{multline}
where the Pochhammer double loop encircles $0$ and $1$. We will now give
three cases of analytic continuations. Other cases can be obtained in a 
similar way. One starts with the still simple case that only one argument
is either close to $1$ or $\infty$. This case can be developed along the
lines set out in \cite{Exton:yx} by rewriting the multiple series in such
a way that the innermost summations is replaced by ordinary Gaussian
hypergeometric series, for which the analytic continuation is known, see eq.\ \ref{contprep} above.
This yields the following results: For the region near infinity, i.e.\ $1/|x_n| < 1$,
the analytic continuation reads
\begin{multline}\label{a:00i}
F_D^{(n)}(a,b_1,\ldots,b_n;c;x_1,\ldots,x_n)\\
 = \frac{\Gamma(c)\Gamma(b_n-a)}{\Gamma(b_n)\Gamma(c-a)}(-x_n)^{-a}
   F_D^{(n)}(a,b_1,\ldots,b_{n-1},1-c+a;1-b_n+a;\frac{x_1}{x_n},\ldots
   \frac{x_{n-1}}{x_n},\frac{1}{x_n})\\
 + \frac{\Gamma(c)\Gamma(a-b_n}{\Gamma(a)\Gamma(c-b_n)}(-x_n)^{-b_n}
   D^{(n)}_{1,1}(a-b_n,b_n,b_1,\ldots,b_{n-1};c-b_n,c-b_n;
   \frac{1}{x_n},x_1,\ldots,x_{n-1}).
\end{multline}
If $|1-x_n|<1$, we are in the region close to $1$, and the analytic 
continuation now reads
\begin{multline}\label{a:001}
F_D^{(n)}(a,b_1,\ldots,b_n;c;x_1,\ldots,x_n)\\
 = \frac{\Gamma(c)\Gamma(c-a-b_n)}{\Gamma(c-a)\Gamma(c-b_n)}
   D^{(n)}_{0,n-1}(a,b_1,\ldots,b_n;a+b_n-c+1,c-b_n;\\
     x_1,\ldots,x_{n-1},1-x_n)\\
 + \frac{\Gamma(c)\Gamma(a+b_n-c)}{\Gamma(a)\Gamma(b_n)}(1-x_n)^{c-a-b_n}
   D^{(n)}_{0,n-1}(c-b_n,b_1,\ldots,b_{n-1},c-a;c-a-b_n+1,c-b_n;\\ 
     x_1,\ldots,x_{n-1},1-x_n).
\end{multline}
The first of the two continuations has the advantage that, in the region of
large $|x_n|$, the first term on the right hand side is convergent even if
all the other $x_i$, $i\neq n$, are close to $1$. Thus, to find the
analytic continuation in the case that one argument is large and another is
close to $1$, one only need to seek the analytic continuation of the 
second term on the right hand side. We will give here the result for the
case that $|x_1|$ is large and $|x_n|$ is close to $1$, since other cases
can easily be obtained by permutations.
\begin{multline}\label{a:i01}
F_D^{(n)}(a,b_1,\ldots,b_n;c;x_1,\ldots,x_n)\\
 = \frac{\Gamma(c)\Gamma(b_1-a)}{\Gamma(b_1)\Gamma(c-a)}(-x_1)^{-a}
   F_D^{(n)}(a,1-c+a,b_2,\ldots,b_{n};1-b_1+a;\frac{1}{x_1},
   \frac{x_2}{x_1},\ldots\frac{x_n}{x_1})\\
 + \frac{\Gamma(c)}{\Gamma(a)}(-x_1)^{-b_1}\Big(
   \frac{\Gamma(c-a-b_n)}{\Gamma(c-b_1-b_n)}
   D^{(n)}_{1,2}(a-b_1,b_1,b_n,b_2,\ldots,b_{n-1};c-b_1-b_n,a+b_n-c+1;\\
   \frac{1}{x_1},1-x_n,x_2,\ldots,x_{n-1})\\
   +\frac{\Gamma(a+b_n-c)}{\Gamma(b_n)}
   (1-x_n)^{c-a-b_n}
   D^{(n)}_{1,2}(c-b_1-b_n,b_1,c-a,b_2,\ldots,b_{n-1};c-b_1-b_n,c-a-b_n+1;\\
   \frac{1}{x_1},1-x_n,x_2,\ldots,x_{n-1})\Big).
\end{multline}

Unfortunately, these formulas are valid only for generic values of the
parameters. In the cases relevant for the Seiberg--Witten periods, we have
certain relations such as $a=b_i$ for some $i$ which will cause 
singularities, if one attempts to analytically continue in the coordinate
$x_i$. To obtain the correct answer, one has to take one further step,
namely a limiting procedure. This is the well known Frobenius process,
which essentially is nothing else than to consider the limit of
$b_i=a+\epsilon$ for $\epsilon\rightarrow 0$. As an example, we present here
one particular instance, namely

\begin{multline}\label{cont}
F_D^{(n)}(a,b_1,\ldots,b_{n-1},a;c;x_1,\ldots,x_n)\\
=\Gamma\!\begin{bmatrix}c\\a,c-a\end{bmatrix}(-x_n)^{-a}
	\sum_M\sum_{m_n=0}^\infty
	\Gamma\!\begin{bmatrix}c-a-\lvert M \rvert\\c-a+\lvert M\rvert \end{bmatrix}\frac{(a)_{\lvert M \rvert+m_n}(1-c+a)_{2\lvert M\rvert+m_n}}
	{(\lvert M \rvert +m_n)!m_n!}\prod_{i=1}^{n-1}\frac{(b_i)_{m_i}}{m_i!}\times\\
	\times\left(\log(-x_n)+h_{m_n}\right)\left(\frac{x_1}{x_n}\right)^{m_1}\cdots
	\left(\frac{x_{n-1}}{x_n}\right)^{m_n-1}\left(\frac{1}{x_n}\right)^{m_n}\\
+\Gamma\!\begin{bmatrix}c,c-a\\a\end{bmatrix}(-x_n)^{-a}\sum_M\sum_{m_n=0}^{\lvert M \rvert-1}
	\frac{(a)_{m_n}\Gamma(\lvert M \rvert - m_n)}{m_n!(c-a)_{\lvert M \rvert-m_n}}
	\prod_{i=1}^{n-1}\frac{(b_i)_{m_i}}{m_i!}x_1^{m_1}\cdots x_{n-1}^{m_{n-1}}\left(\frac{1}{x_n}\right)^{m_n},
\end{multline}
where
\begin{equation}
h_{m_n}=\psi(1+\lvert M \rvert+m_n)+\psi(1+m_n)-\psi(a+\lvert M \rvert+m_n)-\psi(c-a-m_n),
\end{equation}
and we have made use of multindex notation, so that $M=(m_1,\ldots,m_{n-1})$,
$\lvert M \rvert=\sum_{i=1}^{m_n}m_i$ and summation over $M$ means summation over each $m_i$
$(i=1,\ldots,n-1)$ from $0$ to $\infty$.

This result can also be obtained `directly' by using a suitable continuation
formula for the Gaussian hypergeometric function.

Since such limiting procedures make the formulas extremely cumbersome, it
is easier to work with the generic formulas, perform the necessary
expansions with a computer algebra package and to then take the limit.
The Lauricella functions which we encountered in our study of the $SU(2)$
case (cf.\ sect.\ \ref{su2sect}) can be written in the following form: For $a(u)$, one has
$F^{(3)}_D(\frac12,-2,\frac12,\frac12;1;x,\frac{x}{2-x},\frac{x}{2})$, where
$x=1-\frac{e_1}{e_4}$ with the notations used there, i.e.\ $e_1=
\sqrt{u-\Lambda^2}$ and $e_4=\sqrt{u+\Lambda^2}$. For small $x$, this
has a good power series expansion, namely,
\begin{equation}
  F^{(3)}_D(\frac12,-2,\frac12,\frac12;1;x,\frac{x}{2-x},\frac{x}{2})
  = 1 - \frac{3}{4}x + \frac{5}{32}x^2 + \frac{3}{123}x^3 - \frac{169}{262144}
    x^4 - \frac{1131}{1048576}x^5 + \ldots 
\end{equation}
Now, again for small $x$, the dual period is proportional to
$F^{(3)}_D(\frac12,-2,\frac12,\frac12;1;2,2\frac{x-1}{x},2\frac{x-1}{x-2})$
which calls for an analytic continuation valid near the point $(0,\infty,1)$. Actually,
the first argument has modulus greater than $1$, but since the Lauricella function is
only polynomial in its first argument, we do not have to perform an analytic 
continuation for it. We find thus for the dual period
\begin{multline}
  F^{(3)}_D(\frac12,\frac12,-2,\frac12;1;2\frac{x-1}{x},2,2\frac{x-1}{x-2})\\
  = -\frac{\sqrt{2x}}{\pi}\Big(2+3x+\frac{27}{8}x^2+\frac{19}{6}x^3+
     \frac{9559}{4096}x^4 + \frac{12019}{12288}x^5+\ldots\\
  + \log(2)\big(3+\frac{15}{4}x+\frac{135}{32}x^2+\frac{561}{128}x^3
    +\frac{989}{256}x^4+\frac{11169}{4096}x^5+\ldots\big)\\
  - \log(x)\big(1+\frac{5}{4}x+\frac{41}{32}x^2+\frac{147}{128}x^3+
    \frac{193}{256}x^4+\frac{575}{4096}x^5+\ldots\big)\Big).
\end{multline}
Of course, we can also obtain $a_D(u)$ from $a(u)$ by a suitable analytic
continuation. For this, one notes that the Lauricella functions are related to
each other by (for the moment ignoring the fact that the
three variables are all functions of $x=1-\frac{e_1}{e_4}$)
\begin{align}
a(u)&\propto F^{(3)}_D(\frac12,-2,\frac12,\frac12;1;x,y,z),\\
a_D(u)&\propto F^{(3)}_D(\frac12,-2,\frac12,\frac12;1;\frac{x}{z},1-
\frac{1}{y},1-y)\,,
\end{align}
where the prefactors have been omitted. Such relations can be used, for example, to find the 
curve of marginal stability, where $a(u)/a_D(u)$ is real.

\end{document}